\documentclass[preprint,aps,showpacs]{revtex4}
\usepackage{graphicx}

\begin{document}
\title{Nonlinear model for magnetosonic shocklets in plasmas}
\author{P.K.\ Shukla}
\affiliation{Institut f\" ur Theoretische Physik IV, Fakult\"at f\"ur
  Physik und Astronomie, Ruhr--Universit\"at Bochum, D--44780 Bochum,
  Germany} 
\author{B.\ Eliasson}
\affiliation{Institut f\" ur Theoretische Physik IV, Fakult\"at f\"ur
  Physik und Astronomie, Ruhr--Universit\"at Bochum, D--44780 Bochum,
  Germany} 
\author{M.\ Marklund}
\affiliation{Institut f\" ur Theoretische Physik IV, Fakult\"at f\"ur
  Physik und Astronomie, Ruhr--Universit\"at Bochum, D--44780 Bochum,
  Germany} 
\author{R.\ Bingham} 
\altaffiliation[Permanent address: ]{Rutherford Appleton Laboratory,
  Chilton, Didcot, Oxfordshire, OX11 0QX, UK}
\affiliation{Institut f\" ur Theoretische Physik IV, Fakult\"at f\"ur
  Physik und Astronomie, Ruhr--Universit\"at Bochum, D--44780 Bochum,
  Germany} 

\date{Received 20 January 2004}

\begin{abstract}
Exact nonlinear equations for magnetosonic shocklets in a uniform hot 
magnetoplasma are derived by using the nonlinear magnetohydrodynamic
equations. Analytic as well as numerical solutions of the nonlinear
equations are presented.  Shock-like structures of the ion fluid velocity 
and magnetic field (or the plasma density) perturbations are obtained. 
The results may have relevance to the understanding of fast magnetosonic 
shocklets that have been recently observed by onboard instruments of the 
Cluster spacecraft at the Earth's bow shock.
\end{abstract}
\maketitle

In the past, there has been a great deal of theoretical interest 
(e.g. Refs. \cite{r0,r1,r2}) in studying the properties of nonlinear 
magnetohydrodynamic waves in plasmas. It has been found that both slow 
and fast magnetosonic (FMS) waves can propagate  in the form  of
either solitary  or shock waves in plasmas. Very recently 
Stasiewicz {\it et al.} \cite{a0,a1} reported detailed properties 
of slow magnetosonic (SM) solitons \cite{a0} and FMS shocklets \cite{a1}, 
which have been observed by a fleet of four  Cluster spacecraft at the 
quasi-parallel bow shock. Observations reveal that SM solitons \cite{jm} 
are associated with large amplitude compressional (rarefactional)  
magnetic field (plasma density) variations. On the other hand, FMS shocklets 
are accompanied with compression of the plasma density and magnetic field 
perturbations.  

In this Brief Communication, we present a nonlinear model for FMS shocklets, which 
may account for the observed FMS shocklets at the Earth's bow shock. Specifically, 
we show that FMS shocklets are associated with the nonlinear steepening \cite{r3} 
of arbitrary large  amplitude FMS waves in a high-beta magnetoplasma.

The dynamics of the nonlinear FMS waves in a magnetized plasma is governed by 
the inertialess electron momentum equation

\begin{equation}
0 = -n_e e {\bf E} - \nabla p_e -n_e e\frac{{\bf v}_e}{c}\times {\bf B},
\label{eq1}
\end{equation}
where $n_e$ is the electron density, $e$ is the magnitude of the
electron charge, ${\bf E}$ is the wave electric field, ${\bf B}$
is the sum of the ambient and wave magnetic fields, $p_e =n_e T_e$ 
is the electron pressure, $T_e$ is the electron temperature, ${\bf v}_e$
is the electron fluid velocity, and $c$ is the speed of light in  vacuum.
The ion dynamics is governed by the ion continuity equation

\begin{equation}
\frac{\partial n_i}{\partial t} + \nabla \cdot (n_i {\bf v}_i) =0,
\label{e2}
\end{equation}
and the ion momentum equation

\begin{equation}
\rho \left( \frac{\partial}{\partial t}
+ {\bf v}_i \cdot \nabla\right) {\bf v}_i = n_i e {\bf E}-\nabla p_i 
+ n_i e \frac{{\bf v}_i}{c}\times {\bf B},
\label{e3}
\end{equation}
where $n_i$ is the ion number density, ${\bf v}_i$ is the ion fluid velocity, 
$\rho =n_i m_i$ is the ion mass density, $m_i$ is the ion mass,  $p_i =n_i T_i$ 
is the ion pressure, and $T_i$ is the ion temperature. Equations (1)-(3) are 
closed by means of Amp\`ere's and Faraday's laws

\begin{equation}
\nabla \times {\bf B} =\frac{4\pi e}{c}(n_i {\bf v}_i -n_e {\bf v}_e),
\label{e4}
\end{equation}

\begin{equation}
\frac{\partial {\bf B}}{\partial t} =-c \nabla \times {\bf E},
\label{e5}
\end{equation}
together with the quasi-neutrality condition $n_e =n_i=n$. 
The latter is valid for a dense plasma in which the ion plasma
frequency is much larger than the ion gyrofrequency. Equation (4) holds
for the FMS waves whose phase speed is much smaller than the speed of light.

Eliminating ${\bf E}$ from (1) and (3) we obtain

\begin{equation}
\frac{\partial {\bf v}_i}{\partial t} + {\bf v}_i \cdot \nabla {\bf v}_i
 = \frac{(\nabla \times {\bf B}) \times {\bf B}}{4\pi \rho}
 -C_s^2 \nabla {\rm ln }\, \rho,
 \label{e6}
 \end{equation}
where the ion sound speed is denoted by $C_s =[(T_e+T_i)/m_i]^{1/2}$.
On the other hand, from (1), (4) and (5) we have

\begin{equation}
\frac{\partial {\bf B}}{\partial t} =
\nabla \times \left[{\bf v}_i \times {\bf B} 
-\frac{c}{4\pi en}(\nabla \times {\bf B})\times {\bf B}\right].
\label{e7}
\end{equation}

We are interested in studying the nonlinear properties of one-dimensional
FMS waves across the external magnetic field direction $\hat {\bf z} B_0$,
where $\hat {\bf z}$ is the unit vector along the $z$ axis and $B_0$ is the
strength of the ambient magnetic field. Thus, we have $\nabla =\hat {\bf x}
\partial/\partial x$, ${\bf v}_i =u \hat {\bf x}$ and ${\bf B}
=B(x)\hat {\bf z}$, where $\hat {\bf x}$ is the unit vector along the $x$ axis 
in the Cartesian coordinate. Normalizing $\rho$ by the equilibrium mass density $\rho_0
=n_0 m_i$,  $u$ by the Alfv\'en speed $V_A =B_0/\sqrt{4\pi \rho_0}$, $B$ by $B_0$,
time by the ion gyroperiod $\omega_{ci}^{-1}$, $x$ by $V_A/\omega_{ci}=c/\omega_{pi}$,
where $\omega_{pi}$ is the ion plasma frequency, we have our nonlinear MHD equations 
in dimensionless form as

\begin{equation}
\frac{\partial \rho}{\partial t} + \frac{\partial}{\partial x} (\rho u) =0,
\label{e8}
\end{equation}

\begin{equation}
\frac{\partial u}{\partial t} + u \frac{\partial u}{\partial x}
+\frac{1}{\rho} \frac{\partial}{\partial x} \left(\frac{h^2}{2} 
+\beta\,  \rho\right)=0, 
\label{e9}
\end{equation}
and 

\begin{equation}
\frac{\partial h}{\partial t} + h\frac{\partial u}{\partial x}
+ u \frac{\partial h}{\partial x} =0,
\label{e10}
\end{equation}
where $h= B (x,t) /B_0$ and $\beta= C_s^2/V_A^2 \equiv 4\pi n_0(T_e+T_i)/B_0^2$ 
represents the plasma beta.

Equations (8) and (10) yield $\rho=h$, a concept of frozen-in-field lines  
in a magnetized plasma. Hence, we have from (9) and (10)

\begin{equation}
\frac{\partial u}{\partial t} + u\frac{\partial u}{\partial x}
+ \frac{\partial }{\partial x}\left(h + \beta \, {\rm ln}\, h\right)=0,
\label{e11}
\end{equation}
and

\begin{equation}
\frac{\partial h}{\partial t}+ u\frac{\partial h}{\partial x}
+ h \frac{\partial u}{\partial x} =0.
\label{e12}
\end{equation}
In the zero-$\beta$ limit, Eqs. (11) and (12) agree completely with 
Eqs. (2a) and (2b) of Stenflo {\it et al.} \cite{a2} who demonstrated 
rapid steepening of the velocity and magnetic field perturbations 
leading to the formation of FMS shocklets in a cold magnetoplasma.

In the following, we study the properties of FMS shocklets in a warm 
magnetoplasma.  We introduce the change of variables

\begin{eqnarray}
\label{psi1}
\psi_1 &=& u+2\sqrt{h+\beta}+\sqrt{\beta}\mathrm{ln}
\left(\frac{\sqrt{h+\beta}-\sqrt{\beta}}{\sqrt{h+\beta}+\sqrt{\beta}}\right)
+ c_1,
\\
\label{psi2}
\psi_2 &=& -u+2\sqrt{h+\beta}+\sqrt{\beta}\mathrm{ln}
\left(\frac{\sqrt{h+\beta}-\sqrt{\beta}}{\sqrt{h+\beta}+\sqrt{\beta}}\right)
+ c_2,
\end{eqnarray}
where $c_1$ and $c_2$ are constants. This diagonalizes the system of equations 
(\ref{e11})--(\ref{e12}) to the form 

\begin{eqnarray}
\label{psi1eq}
\frac{\partial \psi_1}{\partial t}+(u+\sqrt{h+\beta})
\frac{\partial \psi_1}{\partial x}=0,
\\
\label{psi2eq}
\frac{\partial \psi_2}{\partial t}+(u-\sqrt{h+\beta})
\frac{\partial \psi_2}{\partial x}=0,
\end{eqnarray}
where $u$ and $h$ are given in terms of $\psi_1$ and $\psi_2$
by Eqs. (\ref{psi1})--(\ref{psi2}).  This system admits ``simple wave'' 
solutions \cite{a6}, which can be found by either setting $\psi_1$ or $\psi_2$ 
to zero.  Setting $\psi_2$ to zero in Eqs. (\ref{psi1eq})--(\ref{psi2eq}), we
obtain 

\begin{equation}
\label{steep1}
\frac{\partial \psi_1}{\partial t}+(u+\sqrt{h+\beta})
\frac{\partial \psi_1}{\partial x}=0,
\end{equation}
where from Eqs. (\ref{psi1})--(\ref{psi2}) 

\begin{eqnarray}
\label{h}
u=2\sqrt{h+\beta}+\sqrt{\beta}\mathrm{ln}
\left(\frac{\sqrt{h+\beta}-\sqrt{\beta}}{\sqrt{h+\beta}+\sqrt{\beta}}\right)
 + c =\frac{\psi_1}{2} .
\end{eqnarray}
The constant 
$$c = -2\sqrt{1 + \beta} -
\sqrt{\beta}\ln\left(\frac{\sqrt{1 + \beta} - \sqrt{\beta}}{\sqrt{1 +
 \beta} + \sqrt{\beta}}\right)$$
is determined by the boundary conditions $u = 0$ and $h = 1$ at $|x| =
\infty$.  Using $\psi_1=2u$ in Eq. (\ref{steep1}), we have

\begin{equation}
\label{u}
\frac{\partial u}{\partial t}+(u+\sqrt{h+\beta})
\frac{\partial u}{\partial x}=0,
\end{equation}
where $h(u)$ is implicitly given by Eq. (\ref{h}).
We note that the results given by Eqs. (\ref{h}) and (\ref{u})
generalize the results presented in Ref. \cite{a2} for the
case of arbitrary $\beta$ values.  The paths in the ($x,t$) space 
where $u$ is constant, can be described by the ordinary differential equation

\begin{equation}
\frac{dx}{dt}=u+\sqrt{h+\beta},
\end{equation}
where the right-hand side is constant (since $u$ is constant along
the path), which after integration gives
 
\begin{equation}
x(t)=(u+\sqrt{h+\beta})t+x_0.
\end{equation}
Here $x_0$ is a constant of integration. The general solution of 
Eq. (\ref{u}) is a function of the integration constant $x_0$, viz.

\begin{equation}
  \label{analytic}
  u=f(x_0)=f[x-(u+\sqrt{h+\beta})t],
\end{equation}
where $h(u)$ is given by Eq. (\ref{h}) and $f$ is a function of one variable, 
determined from the initial condition at $t=0$; the velocity $u$ can 
be evaluated for different $x$ and $t$ by solving Eq. (\ref{analytic}) for $u$. 
Equation (\ref{analytic}) describes a nonlinear FMS wave propagating in the 
positive $x$ direction, where the time-dependent  solution has a typical structure 
of the wave-steepening, similarly to the solution of the inviscid Burgers 
equation \cite{r1}. This solution may also be obtained by directly assuming 
that $h$ can be written as a function of $u$. Note that there are no steady state 
solutions within this system, since the dispersion is absent; but by including the 
electron inertial effect in Eq. (\ref{e10}) this could be achieved on
length scales $\lesssim \lambda_e=c/\omega_{pe}$, where $\omega_{pe}$ is the
electron plasma frequency. The dispersive effects break the $\rho =h$ relation, 
and produce an asymmetry between the density and magnetic field perturbations.

We have analyzed the system (\ref{e11})--(\ref{e12}) numerically. 
As an initial condition, we took the magnetic field
$h=1+0.5\,\mathrm{sech}(x/20)$, describing a localized magnetic 
field (and density) compression of the plasma. As we are interested
in the evolution and creation of a ``shocklet'' moving in one 
direction, we choose the initial condition for the velocity 
$u$ from Eq. (\ref{h}), describing a wave moving in the positive $x$--direction 
only. The evolution of the system for  a low-beta ($\beta=0$) and  for a 
high-beta ($\beta=10$) plasma is displayed in Figs. 1 and 2, respectively. 
In agreement with the analytical prediction, the initial wave is propagating 
in the positive $x$--direction only.  Figure 1 shows that both the magnetic 
and velocity perturbations steepen and a shock front starts to develop. 
For the high-beta plasma case, as displayed in Fig. 2, the velocity associated 
with the shocklet is larger, and the self-steepening develops faster than for the 
low-beta plasma case. When the shock fronts become steep enough, effects
such as the electron inertia and electron Landau damping (which heats the plasma) 
will become important. Dispersive effects also occur if the waves propagate
with some angle to the magnetic field direction. The combined effects of 
dispersion and wave-particle induced dissipation could explain the apparent 
phase asymmetry between the magnetic field and density perturbations, as observed 
in large-amplitude FMS shocklets \cite{a1}. This asymmetry is likely to appear 
after the shocklets have developed due to the self-steepening of the FMS waves,  
as investigated here.

To summarize, we have considered the nonlinear propagation of FMS waves in 
a hot magnetoplasma. It has been shown that the nonlinear MHD equations 
in a finite-$\beta$ plasma can be reduced to a pair of equations in which 
the ion fluid velocity and the compressional magnetic field are nonlinearly 
coupled. The system has been diagonalized and special, single wave solutions 
have been obtained. The solutions represent the spatio-temporal evolution 
of an arbitrary amplitude FMS waves. The equations for the full system are 
solved numerically to show the formation of FMS shocklets, in full agreement 
with the analytic results. The finite plasma beta has significant influence 
on the shocklet profile in that the shocks develop on a much shorter timescale 
than for the the plasma with a low beta. In conclusion, the present results 
qualitatively account for the salient features of the observed FMS shocklets 
at the Earth's bow shock \cite{a1}.

\acknowledgments
This work was partially supported by the European Commission (Brussels, Belgium)  
through contract No. HPRN-CT-2000-00314 for carrying out the task of the Human
Potential Research Training Networks ``Turbulent Boundary Layers in
Geospace Plasmas'',  as well as by the Deutsche Forschungsgemeinschaft 
(Bonn, Germany) through the Sonderforschungsbereich 591 entitled ``Universelles 
Verhalten Gleichgewichtsferner Plasmen: Heizung, Transport und Strukturbildung''.

\newpage

\noindent FIGURE CAPTIONS

\bigskip

\noindent
FIG. 1. The evolution of the normalized ion fluid velocity (upper panel) and 
compressional magnetic field (lower panel) for $\beta=0$, for the times 1) 
$t=0$, 2) $t=12.25$ 3) $t=24.75$ and 4) $t=37.25$.

\bigskip\bigskip

\noindent
FIG. 2.  The evolution of the normalized ion fluid velocity (upper panel) and 
compressional magnetic field (lower panel) for $\beta=10$, for the times 1) 
$t=0$, 2) $t=12.25$ and 3) $t=24.75$.

\newpage

\begin{figure}
\includegraphics{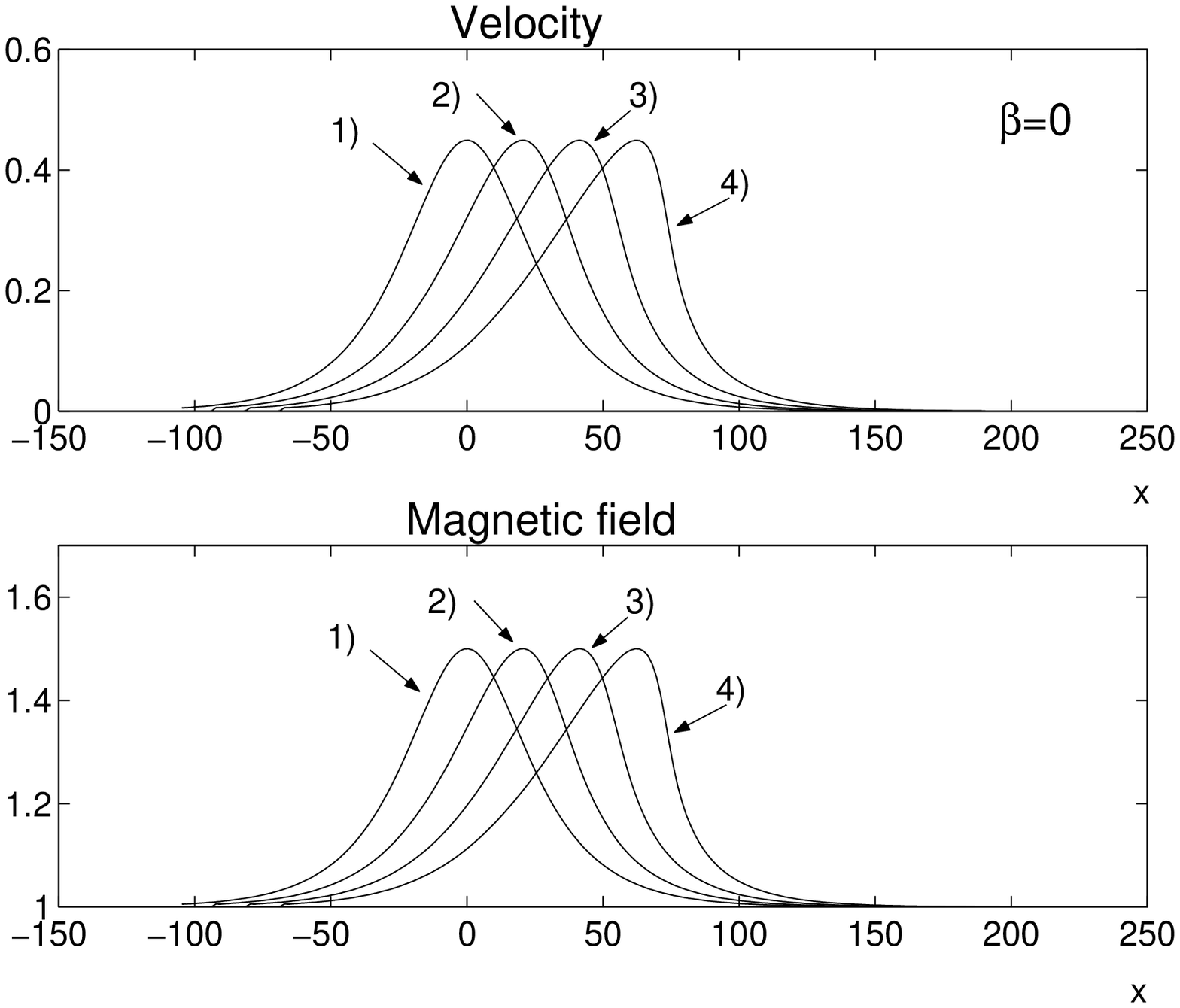}
\caption{}
\end{figure}

\begin{figure}
\includegraphics{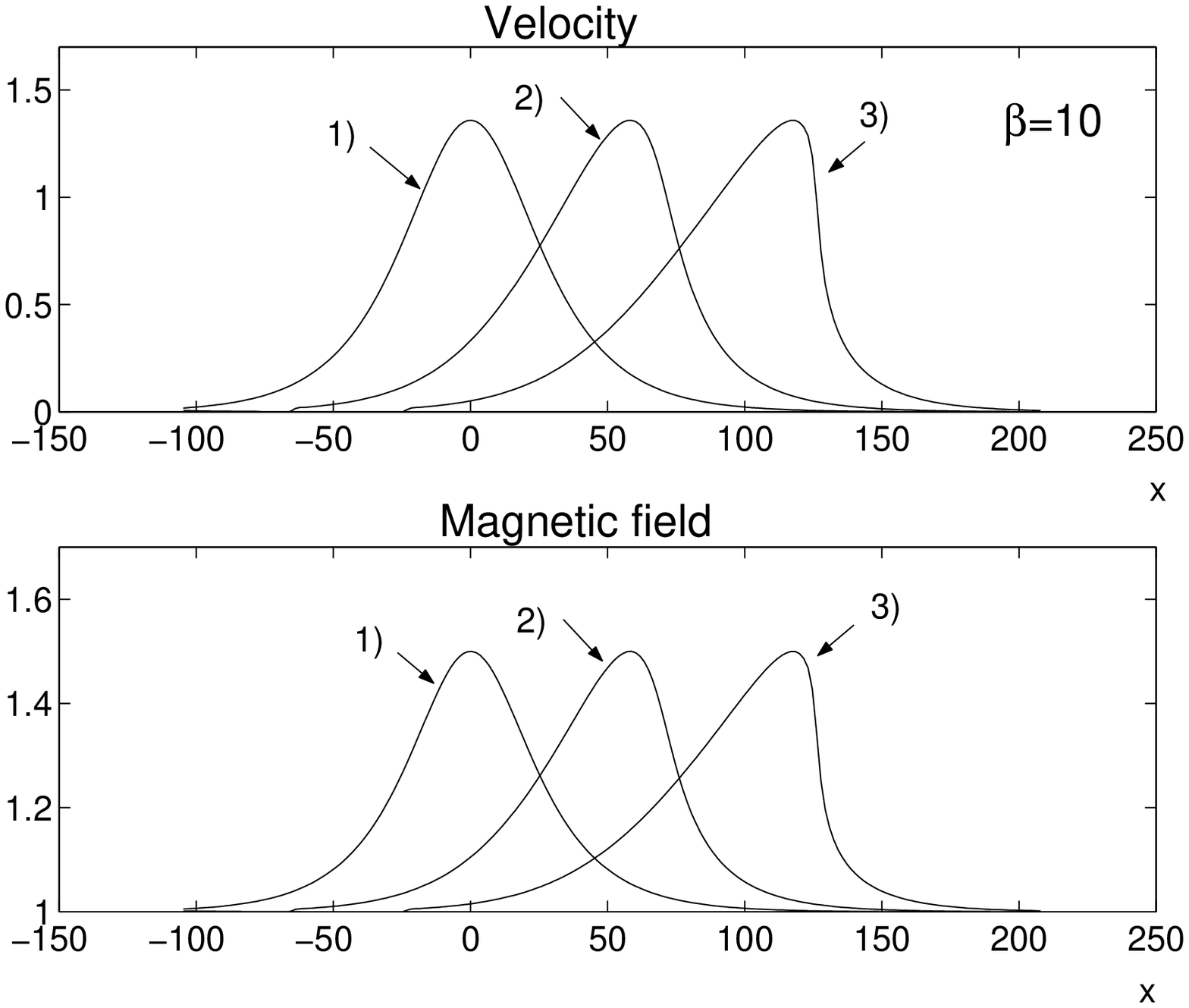}
\caption{}
\end{figure}

\end{document}